\documentclass[twocolumn,showpacs,preprintnumbers,amsmath,amssymb]{revtex4}
\usepackage{amsfonts}

\usepackage{graphicx}
\usepackage{dcolumn}
\usepackage{bm}
\usepackage{latexsym}
\usepackage{float}

\begin{document}


\title{Evolutionary method for finding communities in bipartite networks}

\author{Weihua Zhan$^{1}$}
\email{08zhanwh@tongji.edu.cn}
\author{Zhongzhi Zhang$^{2,3}$}
\email{zhangzz@fudan.edu.cn}
\author{Jihong Guan$^{1}$}
\email{jhguan@tongji.edu.cn}
\author{Shuigeng Zhou$^{2,3}$}
\email{sgzhou@fudan.edu.cn}

\affiliation{$^{1}$Department of Computer Science and Technology,
Tongji University, 4800 Cao'an Road, Shanghai 201804, China}

\affiliation {$^{2}$School of Computer Science, Fudan University,
Shanghai 200433, China}

\affiliation {$^{3}$Shanghai Key Lab of Intelligent Information
Processing, Fudan University, Shanghai 200433, China}

\date{\today}

\begin{abstract}
An important step in unveiling the relation between network
structure and dynamics defined on networks is to detect communities,
and numerous methods have been developed separately to identify
community structure in different classes of networks, such as
unipartite networks, bipartite networks, and directed networks.
Here, we show that the finding of communities in such networks can be
unified in a general framework--- detection of community structure in
bipartite networks. Moreover, we propose an evolutionary method for
efficiently identifying communities in bipartite networks. To this
end, we show that both unipartite and directed networks can
be represented as bipartite networks, and their modularity is
completely consistent with that for bipartite networks, the
detection of modular structure on which can be reformulated as
modularity maximization. To optimize the bipartite modularity, we
develop a modified adaptive genetic algorithm (MAGA), which is shown
to be especially efficient for community structure detection. The high
efficiency of the MAGA is based on the following three improvements
we make. First, we introduce a different measure for the informativeness
of a locus instead of the standard deviation, which can exactly
determine which loci mutate. This measure is the bias between the
distribution of a locus over the current population and the uniform
distribution of the locus, i.e., the Kullback-Leibler divergence between
them. Second, we develop a reassignment technique for
differentiating the informative state a locus has attained from the
random state in the initial phase. Third, we present a modified
mutation rule which by incorporating related operation can guarantee
the convergence of the MAGA to the global optimum and can speed up the
convergence process. Experimental results show that the MAGA outperforms
existing methods in terms of modularity for both bipartite and
unipartite networks.
\end{abstract}

\maketitle

\section{Introduction}
Complex network has gained overwhelming popularity as a powerful
tool for understanding various complex systems from diverse fields,
including the technical, natural, and social sciences, etc., which
provides a unified perspective or method for studying these systems
through modeling them as networks with nodes and edges respectively
representing their units and interactions between
units~\cite{Watts98,Barabasi99,Albert02,Dorogo03,Newman03,Boccaletti06}.
Generally, according to the types of node, networks can be
classified into unipartite, bipartite and multipartite networks. As
a typical class of real-world networks, bipartite networks, compared
to unipartite ones, consist of two types of nodes, and edges exist only
 between distinct types of nodes. Examples of bipartite
networks come from various fields, including scientific
collaboration networks, actor-movie networks, and protein-protein
interaction
networks~\cite{Watts98,Barabasi99,Girvan02,Newman04a,Newman04b}.
Multipartite networks with more than three types of nodes, are
occasionally seen~\cite{Neubauer09,Murata10}.

It has been discovered~\cite{Girvan02} that most real networks
share a local clustering feature, i.e., groups of tight-knit nodes
mutually connected to each other with sparser edges. These groups of
nodes are generally referred to as communities or modules. From a
topological point of view, a community may correspond to a functional
unit because of its relative structural independence. In turn,
community structure can critically affect diverse dynamics on
networks. Therefore, identification of communities plays a key role
in numerous related areas of complex networks, e.g., predicting
protein function~\cite{Alexei03} and determining dynamics of
systems~\cite{Anderson89,Arenas06,Yan07}. The last few years have
witnessed tremendous efforts in this
direction~\cite{Newman04a,Guimera05,Newman06a,Newman06b,Duch05,Arenas06,Zhou03,Radicchi04,Palla05,Guimera07,Leicht08,Freeman03,Barbe07}
(useful reviews include Refs.~\cite{Danon05,Fortunato10}). Most previous studies are dedicated to deal with unipartite networks,
while little attention has been paid to directed
networks~\cite{Guimera07,Leicht08} and bipartite
networks~\cite{Freeman03,Guimera07,Barbe07}.

It is of interest that unipartite and directed networks can
be represented by bipartite networks as will be shown. Thus,
detection of communities in unipartite networks or in directed networks
can be transformed into the same task in bipartite networks. Given a
bipartite modularity, those methods based on modularity
maximization~\cite{Guimera05,Newman06a,Newman06b,Duch05}, in
principle, can be applied to bipartite networks. However, they are
expected to be affected by the resolution
limit~\cite{Fortunato07,Kumpula07} as in the unipartite case, which
may result in the degeneracy problem~\cite{Good10}. This poses a
challenge for the methods that return one solution. Instead, we
present a modified adaptive genetic algorithm to optimize the
bipartite modularity~\cite{Barbe07}. The evolutionary method can
return a better solution in a shorter time. Moreover, the method
also can return multiple better solutions in multiple runs, which
enables us to evaluate the reliability~(or
significance) of solutions without resorting to other technique as
in ~\cite{LaRaRa10,Pardo07}, as well as to obtain a superior
solution by combining several better solutions when the degeneracy
problem is severe.

In practice, there exist two distinct conceptual understandings of
the community structure of a bipartite network. The first viewpoint
for communities in the network is to consider each composed of two
types of nodes with dense edges across them, which is similar to
the view of unipartite cases~\cite{Barbe07}. An alternative view is that
any community should contain only one type of nodes, which are
closely connected through co-participation in several communities
that consist of another type of nodes~\cite{Guimera07}. Guided by
this view, the usual approach to identifying communities is to
project the bipartite network onto one specific unipartite network
as needed, and then identify communities in the projection.
Guimer{\`a} \emph{et al.}~\cite{Guimera07} recently presented a method for
identifying communities of one type of nodes against the other type
of nodes with a known community structure.

In this paper, we focus on dealing with the problem of identifying
communities from the first viewpoint. We present a modified adaptive
genetic algorithm (MAGA), based on the mutation-only genetic
algorithm~(MOGA), which is parameter-free unlike the traditional
genetic algorithms. The method has no need to know in advance the
number of communities and their sizes. In Sec.
\ref{sec:modularity}, we first give a short review of Barber's
modularity~\cite{Barbe07} and then show that unpartite networks and directed
networks can be uniformly represented by bipartite networks. After
the description of the MOGA in Sec. \ref{subsec:moga}, we introduce a
different measure for selecting loci to mutate in Sec. \ref{subsec:newmeasure},
and then develop the reassignment technique in
Sec.~\ref{subsec:reassignment}. Further, we discuss how to select the
population size in Sec.~\ref{subsec:popusize} and address the issues of
convergence and time complexity of the MAGA in Sec.~\ref{subsec:covergence}.
In Sec.~\ref{sec:results}, we apply the algorithm to model
bipartite networks, several real bipartite networks, and unipartite
networks. Finally, the conclusion is given.

\section{Bipartite modularity}\label{sec:modularity}

The modularity introduced by Newman and Girvan~\cite{Newman04a} aims at
quantifying the goodness of a particular division of a given
network, and has been widely accepted as a benchmark index to
measure and to compare the accuracy of various methods of community
detection. The definition of this quantity is based on the idea that
community structure definitely means a statistically surprising
arrangement of edges, that is, the number of actual edges within
communities should be significantly beyond that of the expected edges of
a null model. In turn, a null model should have the same number of
nodes and degree distribution as the original network, while the edges
of the null model are placed by chance.

Let $k_i$ be the degree of nodes $i$, and $M$ the total number of edges.
Since in the null model~\cite{Newman06b} the probability for an edge
being present between nodes $i$ and  $j$ is $\frac{k_ik_j}{2M}$, the
modularity quantifying the extent of the number of actual edges exceeding the expectation based on the null model network, can be formulated as follows:
\begin{equation}\label{eqmodularity}
Q=\frac{1}{2M}\sum^N_{i=1}\sum^N_{j=1}\left(A_{i,j}-\frac{k_ik_j}{2M}\right)\delta(g_i,g_j)
\end{equation}
where $Q$ is the sum of the difference over all groups of the
particular division, $N$ is the network size, $A_{i,j}$ indicates
the adjacent relation between nodes $i$ and $j$, $g_i$ represents
the group the node~\emph{i} is assigned to, and the $\delta$
function takes the value of 1 if $g_i$ equals $g_j$, 0 otherwise.

The value of $Q$ ranges from -1 to 1. Given a network, a larger
value generally indicates a more accurate division of the network
into communities. Community structure detection thus can be
formulated as a problem of modularity maximization, which often
works well although it may suffer from a resolution
problem~\cite{Fortunato07,Kumpula07}. But on the other hand, due to
the resolution limitation and the random fluctuation
effect~\cite{Guimera04}, it appears preferable for the divisions
delivered by maximization modularity approaches to give an
evaluation of their
reliability~\cite{Karrer08,Good10,LaRaRa10,Pardo07}.

The above modularity is actually designed for unipartite networks.
To be suitable for various networks, several variations of
modularity based on different null models have been proposed,
including weighted~\cite{Arenas07}, directed~\cite{Leicht08}, and
bipartite modularity~\cite{Guimera07,Barbe07}. A bipartite network
with $N$ nodes can be conveniently denoted by a duality
$(p,q)~(p+q=N)$, where $p$ and $q$ respectively represent the
numbers of the two types of nodes. We can renumber nodes such that
in the sequence $1,2,\cdots,p,p+1,\cdots,N$, the leftmost $p$
indices represent the first type of nodes and the remainder
represent the second type of nodes. Then, Barber's bipartite
modularity~\cite{Barbe07}, which considers a community composed of
distinct types of nodes in the network, can be written as
\begin{equation}\label{eqbimodularity}
Q_b=\frac{1}{M}\sum^p_{i=1}\sum^{N}_{j=p+1}\left(A_{i,j}-\frac{k_ik_j}{M}\right)\delta(g_i,g_j)
\end{equation}

\begin{figure}
\includegraphics[scale=0.55]{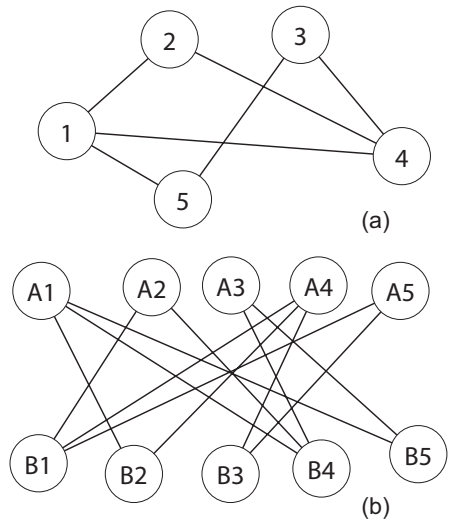}
\caption{Transformation of a simple unipartite network into
a bipartite one. (a) An unipartite network with five nodes and six edges.
(b) The bipartite network corresponding to (a).}
\label{fig_transformation}
\end{figure}

Immediately, a subtle difference between the two modularities in
Eqs.~(\ref{eqmodularity}) and~(\ref{eqbimodularity}) can be
observed. It is of interest that a unipartite network can be
equivalently represented as a bipartite one, and the bipartite
modularity can recover the modularity for the original network. If
each node $i$ is represented by two nodes $A_i$ and $B_i$ and each edge
$i$-$j$ represented by two edges $A_i$-$B_j$ and $A_j$-$B_i$, then a
unipartite network with $N$ nodes and $M$ edges is transformed into
a corresponding bipartite network with $2N$ nodes and $2M$ edges.
For example, the transformation of a simple unipartite network is
shown in Fig. \ref{fig_transformation}.
Further, if we label $N$ nodes $A_i$ with $1, 2, \ldots,N$ and label
$B_i$ with $N+1, N+2, \ldots, 2N$, then an edge $i$-$j$ in the original
network corresponds to two edges, $i$-$(N+j)$ and $j$-$(N+i)$. Using the
bipartite modularity introduced in Eq.~(\ref{eqbimodularity}) on the
induced bipartite network, we have
\begin{eqnarray}
Q_b &=&\frac{1}{2M}\sum^N_{i=1}\sum^{2N}_{j=N+1}\left(\tilde{A}_{i,j}-\frac{k_ik_j}{2M}\right)\delta(g_i,g_j)\nonumber\\
&=&\frac{1}{2M}\sum^N_{i=1}\sum^{N}_{j'=1}\left(\tilde{A}_{i,N+j'}-\frac{k_ik_{N+j'}}{2M}\right)\delta(g_i,g_{N+j'})\nonumber\\
&=&\frac{1}{2M}\sum^N_{i=1}\sum^{N}_{j=1}\left(A_{i,j}-\frac{k_ik_{j}}{2M}\right)\delta(g_i,g_{j})=Q
\end{eqnarray}
where we have made use of the fact that the node $A_i$ and $B_i$ should
be in an identical community and have the same degree. Thus,
bipartite modularity can also be used to community detection in
unipartite networks after being transformed.

We then turn to the modularity for directed unipartite networks, which
are another important class of networks. The directed network can
analogously be transformed to a bipartite network. A node $i$ is
represented by two nodes $A_i$ and $B_i$ as in unipartite
networks, while a directed edge from $i$ to $j$ is represented as an
edge between $A_i$ and $B_j$, that is, set $\{A_i\}$ and set
$\{B_i\}$ are the sources and the sinks. Again, using the
Eq.~(\ref{eqbimodularity}) and the fact above, we obtain
\begin{eqnarray}
Q_b &=&\frac{1}{M}\sum^N_{i=1}\sum^{2N}_{j=N+1}\left(\tilde{A}_{i,j}-\frac{k_ik_j}{M}\right)\delta(g_i,g_j)\nonumber\\
&=&\frac{1}{M}\sum^N_{i=1}\sum^{N}_{j'=1}\left(\tilde{A}_{i,N+j'}-\frac{k_ik_{N+j'}}{M}\right)\delta(g_i,g_{N+j'})\nonumber\\
&=&\frac{1}{M}\sum^N_{i=1}\sum^{N}_{j=1}\left(A_{i,j}-\frac{k^{\rm
out}_ik^{\rm in}_{j}}{M}\right)\delta(g_i,g_{j})
\end{eqnarray}
where the term on right-hand side in the last equation is just the
modularity for directed networks presented in~\cite{Leicht08}. The
method for transforming directed networks into bipartite ones has
been proposed by Guimer{\`a} \emph{et al}~\cite{Guimera07}, but their
bipartite modularity is distinct from Barber's, as mentioned before.

Consequently, the bipartite network can be considered as a wider
class of networks that provides a generic case for the problem of
community structure detection. And Barber's bipartite modularity
can served as a uniform objective for these methods of
identifying communities based on optimization.

\section{Evolutionary method for community detection}\label{sec:evolution}

As a class of general-purpose tools to solve various hard problems,
genetic algorithms have found wide application in bioinformatics,
computer science, physics, engineering, and other fields. They are,
based on the Darwinian principle of survival of the fittest, a kind
of global optimization method simulating evolutionary processes of
species in nature~\cite{Holland75}.

The evolutionary methods are easy to implement, and the process
can be described as follows. The methods start with a stochastically
created initial population with predefined size wherein individuals
are known as chromosomes representing a set of feasible solutions to
the problem at hand, with each associated with a fitness value. Then
chromosomes are selected in proportion to their corresponding fitness so that
those fitter individuals would will have multiple copies
and less fit will be discarded in the new population. Next,
genetic operators such as crossover and mutation are performed
according to the respective specified ratios on the population. After
these operations, the population of the next generation has been
reproduced. The above process is iterated to evolve the current
population toward better offspring until the termination criterion
is met.

Since the number of divisions on any given network grows at least
exponentially in the network size, the optimization of modularity is
clearly an NP-hard problem that has been given a rigid proof
in~\cite{Brandes08}, which has motivated an array of heuristic
methods including greedy agglomeration~\cite{Newman04b}, simulated
annealing (SA)~\cite{Guimera05}, spectral relaxation
(SR)~\cite{Newman06a,Newman06b}, extremal
optimization~(EO)~\cite{Duch05} and mathematical
programming~\cite{Agarwal08}. All these methods perform a point-point
search, that is, transformation from one solution to a better one,
and are susceptible to trapping in a local optimum. In contrast,
genetic algorithms work with a population of solutions instead of a
single solution. This implies that genetic algorithms are more
robust because they perform concurrent searches in multiple directions which would
make them effectively find better solutions.

However, for practitioners, a fundamental important problem is to
choose appropriate parameters such as crossover rate and mutation
rate, because they will seriously affect the performance of genetic
algorithms. Furthermore, these parameters are closely related to the
studied problems, and even for the same problem, they should adjust
themselves in the course of the search. In the following, we would
like to introduce an adaptive genetic algorithm recently presented
by Szeto and Zhang~\cite{Szeto06} and then propose a modified
version suited for community structure detection.

\subsection{Mutation only Genetic algorithm}\label{subsec:moga}

Traditional genetic algorithms assume that genetic operators
indiscriminately act on each locus constituting the chromosome, but
this is not always the case. Indeed, the recent research in human
DNA~\cite{Brinkmann98} shows that mutation rates at different loci are
very different from one another. Inspired by this, Ma and
Szeto~\cite{Ma04} reported on a locus-oriented adaptive genetic
algorithm~(LOAGA) that makes use of the statistical information
inside the population to tune the mutation rate at an individual locus.
Szeto and Zhang~\cite{Szeto06} further presented a new adaptive
genetic algorithm, called the mutation only genetic algorithm (MOGA),
which generalized the LOAGA by incorporating the information about
the loci statistics in the mutation operator. In the MOGA, mutation is
the only genetic operator, and the only required parameter is the
population size. The MOGA was readdressed by Law and Szeto
in~\cite{Law07}, wherein it was extended to include a crossover
operator. Here, the description for the MOGA is given on the basis
of the later version.

The population matrix $P$ has $N_P$ stacked chromosomes with
length $L$, with its entries $P_{ij}(t)$ representing the allele at
locus $j$ of the chromosome $i$ at time (or generation) $t$. The
rows of this matrix are ranked according to the fitness of the
chromosomes in descending order, i.e., $f(i)\ge f(k)$ for $i<k$. The
columns are ranked according to the standard deviation
$\sigma_t(j)$~(its definition will given below) of alleles at locus
$j$ such that $\sigma_t(j)\ge \sigma_t(k)$ for $j<k$. In the MOGA,
the fitness cumulative probability, as an informative measure for
chromosome $i$ relative to the landscape of fitness of the whole
population, was introduced and defined as
\begin{equation}\label{eq:cumufitness}
C(i)=\frac{1}{N_P}\sum_{g\le f(i)}N(g)\,,
\end{equation}
where $N(g)$ is the number of chromosomes whose fitness values equal
$g$. Subsequently, the standard deviation $\sigma_t(j)$ over the
allele distribution, as a useful informative measure for each locus
$j$, is defined as
\begin{equation}\label{eq:deviation}
\sigma_j(t)=\sqrt{\frac{\sum^{N_P}_{i=1}(P_{ij}(t)-\overline{h_j(t)})^2\times{C(i)}}{\sum^{N_P}_{i=1}C(i)}}\,,
\end{equation}
where the weighting factor $C(i)$ reflects the informative usefulness
of the chromosome $i$, and $\overline{h_j(t)}$ is the mean of the
alleles at locus $j$, given by
\begin{equation}
\overline{h_j(t)}=\frac{1}{N_P}\sum^{N_P}_{i=1}P_{ij}(t)\,.
\end{equation}

A locus with a smaller allele standard deviation was considered to be
more informative than other loci, and vice versa. Indeed, this
really makes sense in limited situations. For the initial
population, the alleles at each locus $j$ should satisfy a uniform
distribution, so the standard deviation $\sigma_t(j)$ will be very
high while the locus present is not informative. A typical
optimization problem generally allows for a few global optima, so the
loci with higher structural information are liable to take fewer
alleles than allowed, thereby having smaller allele standard
deviations. Therefore, the loci with higher deviations prefer
mutating while the other loci~(informative loci) remain to guide
the evolution process.

Now we can describe the process for the MOGA. In each generation,
we sweep the population matrix from top to bottom. Each row~(a
chromosome) is selected for mutation, with probability
$\alpha(i)=1-C(i)$. According to Eq.~(\ref{eq:cumufitness}), we have
$\frac{1}{N_P}\le C(i)\le 1$. Then, a chromosome with a higher fitness
value has fewer chances to be selected, and vice versa. In Particular, the
first chromosome that has the highest fitness value will never be
selected for mutation, while the last one will almost always undergo
mutation for a large enough $N_P$, if $N_P$ normally takes a value
from 50 to 100 as De Jong suggesed~\cite{DeJong75}, for example,
$\alpha(N_P)=1-\frac{1}{N_P}=0.98$ for $N_P=50$. If the current
chromosome selected is $i$, then the number $N(i)$ of loci for mutation
is prescribed as $N(i)=\alpha(i)\times{L}$. Thus, a selected
chromosome with a higher fitness value has fewer loci to mutate,
so that most of the informative loci remain; while a selected
chromosome with a lower fitness value has more less-informative loci to
mutate. In practice, we can mutate the $N(i)$ leftmost loci because
they are less informative than others according to the above
arrangement of loci.

Overall, the MOGA was expected to have a two-fold advantage over
traditional genetic algorithms: first, because there is no need to input parameters
except the population size it can be more available for solving
various problems; second, the mechanism of adaptively adjusting
parameters can make it more effectively perform and obtain better
solutions if it works as expected.

\subsection{Measure for the informativeness of loci}\label{subsec:newmeasure}

Despite these possible advantages, the MOGA cannot be directly
applied to community structure detection due to a drawback that
will be shown. Instead, we present a modified version of the MOGA,
i.e., the MAGA, which is especially suited for the problem of community
structure detection. We note that genetic algorithms have been
applied to this problem in~\cite{Tasgin07,Pizzuti08}, but these
applications are based on standard genetic algorithms~(SGAs).

We begin with the encoding schema of the genetic algorithm for
finding communities in a bipartite network. A useful representation
is the locus-based adjacency representation presented by Park and
Song in~\cite{Park98} where it was used in clustering data and also
has been used for community detection~\cite{Pizzuti08}. In this
encoding schema, a chromosome consists of $N$ loci with a locus for
a node in the network, and the allele at a locus $j$ is the label of
one neighbor of node $j$ in the network. In this way, a chromosome
actually induces a graph that is often disconnected because of the
reduction in connectivity relative to the original network. Given the
connectivity of the community, decoding the division from
a chromosome then amounts to finding all the connected components of
the induced graph. For simplicity, we also call them the connected
components of the chromosome.

\begin{figure}
\includegraphics[scale=0.5]{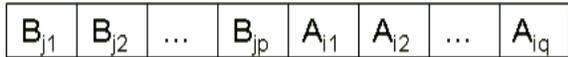}
\caption{Encoding schema of a chromosome for a bipartite network
$(p,q)$. $B_{j_k}$ (for ${k\le p}$) is the allele at the locus
representing node $A_k$, which stands for a neighbor node of $A_k$
in the network. Similarly, $A_{i_k}$ (for $k\le q$) is the allele at
the locus representing node $B_k$.} \label{fig:chromosome}
\end{figure}

Now, we apply the encoding schema to the case of bipartite networks.
Given a bipartite network $(p,q)$, we label its nodes as
noted above, i.e., we label nodes of type $A$ with $1,2,\cdots,p$
while we label another type of nodes with $p+1,\cdots,N$. Then a
chromosome $R$ for the network can be represented as that shown in
Fig.~\ref{fig:chromosome}. Since our objective is to find a division
with as higher a modularity as possible, the fitness function can be defined
directly in terms of the modularity. Based on the above representation
for the chromosome, this function becomes
\begin{equation}\label{eq:fitness}
f({R})=Q_b(\pi_{R})=\frac{1}{M}\sum^p_{i=1}\sum^{N}_{j=p+1}\left(A_{i,j}-\frac{k_ik_j}{M}\right)\delta(g_i,g_j)\,,
\end{equation}
where the parameter $\pi_{R}$ emphasizes that the division on which the
modularity is calculated is encoded by
chromosome $R$.

Recall that in the MOGA the allele standard deviation is used to
pick the loci to mutate. When applied to community structure
detection, however, the measure generally will misguide the
algorithm. Consider the simple case in which the population
consists only of three chromosomes, $R_1$, $R_2$, and $R_3$, which in
turn consist of four loci that have three alleles.
Table~\ref{tab:example} shows the allele distribution at these loci.

From Eqs. (\ref{eq:cumufitness}) and (\ref{eq:deviation}), the
allele standard deviations for the four loci,
$\sigma_1,\sigma_2,\sigma_3$ and $\sigma_4$~(henceforth, we omit the
parameter t for simplicity), can be calculated to obtain
\begin{equation}
\sigma_1>\sigma_2>\sigma_4>\sigma_3.
\end{equation}
According to the selection criterion of the loci to mutate in the MOGA, $\sigma_1$ has the
highest standard deviation and will be picked out.

\begin{table}
\caption{Example of a population with three chromosomes. Fitness is
calculated on the division induced from decoding the chromosome.
Values in each column are the alleles at the locus.}
\label{tab:example}
\begin{ruledtabular}
\begin{tabular}{lddddd}
    Chro. & \multicolumn{1}{c}{Fitness} & \multicolumn{1}{c}{Loc.1} & \multicolumn{1}{c}{Loc.2} & \multicolumn{1}{c}{Loc.3} & \multicolumn{1}{c}{Loc.4} \\ \hline
    $R_1$ & 0.5 & 100 &20 &4 &8\\
    $R_2$ & 0.3 & 100 &50 &5 &12\\
    $R_3$ & 0.2 & 10  &50 &6 &7\\\hline
    $\sigma$ &  & 36.7243  & 15.8114  & 0.8165 &2.0412\\
\end{tabular}
\end{ruledtabular}
\end{table}

In fact, the informativeness of a locus implies a certain bias, and vice
versa. The initial population is generated randomly and each locus
follows an approximately random distribution. From the uniform
distribution, we have nothing on the structure of the optimal solution
to the given problem. With gradual evolution, more and more
fit members of the population will assume the same alleles at some
loci, which may suggest some structural information of the optimal
solutions; that is, the bias~(or deviation) from the random distribution
indicates the informativeness of the locus. In the simplest case
such as the knapsack problem where each locus takes the value 1 or 0, the allele
standard deviation amounts to the bias and the MOGA can work well
\cite{Szeto06}.

For the current case, loci 3 and 4 should be selected with equally
higher priority because their allele distributions are equally
closer to their respective random distributions. Both loci 1 and 2
appear with a certain bias on their alleles, indicating that they are more
informative than others. If the informativeness of each chromosome
is taken into account, however, they are evidently different from one
another. Locus 1 has a larger bias since the chromosomes with the
same allele 100, i.e., $R_1$ and $R_2$, have higher fitness. In
contrast, locus 2 has a smaller bias since the chromosomes with the
same allele 50, i.e., $R_2$ and $R_3$, have lower fitness. Therefore,
the correct order of mutation is
\begin{equation}\label{eq:correctorder}
\text{locus 3=locus 4}>\text{locus 2}>\text{locus 1}\,,
\end{equation}
where the equality means that the pair of loci have the same
priority for mutation. Obviously, the allele standard deviation would
severely misguide the MOGA in the current case.

The failure of the allele standard deviation stems from the fact
that this measure is closely related to alleles at loci. However,
the information contained in loci is actually not relevant to the
particular values but solely determined by the bias relative to the
random distribution. The method of measurement of the bias is thus crucial.
Fortunately, we can use the Kullback-Leibler
divergence~\cite{Kullback51} to describe the bias.

In the formalism of the MAGA, we explicitly represent a locus $j$ as
a discrete random variable $X_j$, and an allele at the locus is
a value that $X_j$ can take. Note that in the following the set of all alleles
at the locus is denoted by $X_j$ as well. Then the random distribution at the
locus can be formally given by
\begin{equation}
\mathcal{Q}(X_j=x)=\begin{cases}
 \frac{1}{|X_j|},
&\text{for each x}\in X_j,\\
0,&\text{otherwise.}
\end{cases}
\end{equation}
Let the allele distribution over the population be $\mathcal {P}$,
defined by
\begin{equation}\label{eq:allele-distribution}
\mathcal{P}(X_j=x)=\frac{\sum_{P_{ij}=x}f(i)}{\sum_if(i)}.
\end{equation}
We can mathematically define the bias $\mu$ as the Kullback-Leibler
divergence between the two distributions,  $\mathcal {P}$ and
$\mathcal {Q}$:
\begin{equation}\label{eq:bias}
\mu(j)=\sum_{x\in X_j}\mathcal{P}(X_j=x)\log
\frac{\mathcal{P}(X_j=x)}{\mathcal{Q}(X_j=x)},
\end{equation}
The base of log is irrelevant, but it will change the value of bias, and in the following all the logs are taken to base 2.
It is noteworthy that the quantity 0log0 should be
interpreted as zero. As a Kullback-Leibler divergence, the bias is
always non-negative and is zero if and only if
$\mathcal{P}=\mathcal{Q}$. The intuitive explanation is that the
amount of information a locus contains is always non-negative, and
that we have to roll an unbiased dice if we have not any knowledge
about something. Conversely, we can predict that an event will
inevitably occur only when we have complete information about it.

Reconsidering the above example, we obtain $\mu_1=0.863$,
$\mu_2=0.585$, and $\mu_3=\mu_4=0.5145$. As a smaller bias indicates
poorer information a locus contains, the locus should undergo
mutation. Conversely, a larger bias means richer informativeness,
and the locus should remain. Therefore, guided by the bias, the order of
mutation is locus 3,4,2,1 or 4,3,2,1, which completely match the
order in Eq.~(\ref{eq:correctorder}).

Furthermore, it can be observed that locus 2 has zero bias if it has
only two alleles. The difference coming from the change of number of
alleles would be normally concealed by the allele standard
deviation. For these reasons, the bias appears superior to the
allele standard deviation. A better alternative is to use the
normalization of the bias as in our MAGA, which ranges from 0 to 1
being divided by $\log |X_j|$.

\subsection{The reassignment technique for the locus
statistic}\label{subsec:reassignment}

It is so far acknowledged that the loci with random distributions
should have the highest priority for mutation. However, in the
community detection case this presupposition does not always hold.
After the evolution of a certain number of generations, some
communities or their main bodies will appear at the population
scale. At present, a locus with a random distribution does not
necessarily imply that it contains no information and should undergo
mutation immediately. Generally, there exist in the network many
nodes whose neighbors are all~(or almost all) in the same communities
and have a similar connection pattern or even are structurally
equivalent nodes~\cite{Lorrain71} that are connected to the same
nodes. For such a node, if all~(or most) of its neighbors presenting
in the same connected component predominates in the current
population, then the locus has a random distribution or an
approximately random distribution. Therefore, we are required to
differentiate the cases to avoid such misguiding.

The reassignment technique is designed to deal with this problem.
For a chromosome $R$, the element $x$ is the allele at the locus $j$
which is a neighbor of the node $j$. Check whether the component
in which $j$ lies includes other neighbors with smaller
labels in the original network. If it is true and the neighbor with the smallest label is
$y$, then the contribution from $R$, $\frac{f(R)}{\sum_if(i)}$ that
should be assigned to $x$ now is reassigned to $y$ if $x\neq y$. In this
way, forward sweeping of the population matrix can obtain an
updated allele distribution at the locus over the population, given by
\begin{equation}\label{eq:allele-newdistr}
\mathcal{P}^*(X_j=x)=\frac{\sum_{\mathcal{S}(i,j)=x}f(i)}{\sum_if(i)}.
\end{equation}
where $\mathcal{S}(i,j)$ is the node $j$'s neighbor with the
smallest label that lies with $j$ in the same component of the
chromosome $i$.

\begin{table}[h]
\caption{Example of reassignment technique. Column 1 lists four
chromosomes, column 2 is the fitness of the chromosomes, column 3 shows the
alleles of locus 1, and the right four columns show whether the
corresponding nodes are in the same connected component as node 1,
with 1 indicating yes and 0 no. } \label{tab:reassign}

\begin{ruledtabular}
\begin{tabular}{ldddddd}
    Chro. & \multicolumn{1}{c}{Fitness} & \multicolumn{1}{c}{Loc.1} & \multicolumn{1}{c}{Loc.2} & \multicolumn{1}{c}{Loc.3} & \multicolumn{1}{c}{Loc.4} & \multicolumn{1}{c}{Loc.5}\\ \hline
    R1 & 0.28 & 2 &1 &1 &0 &0\\
    R2 & 0.25 & 3 &0 &1 &1 &0\\
    R3 & 0.25 & 4 &1 &0 &1 &1\\
    R4 & 0.22 & 5 &0 &1 &0 &1\\
\end{tabular}
\end{ruledtabular}
\end{table}

An example using the technique as shown in Table \ref{tab:reassign}.
Using Eq.~(\ref{eq:allele-distribution}), it is obvious that the
locus 1 has an approximately random distribution and thus the bias
is close to 0. Recalculating the distribution with the reassignment
technique, however, we have $\mathcal{P}^*(X_1=2)=0.53$,
$\mathcal{P}^*(X_1=3)=0.47$, and
$\mathcal{P}^*(X_1=4)=\mathcal{P}^*(X_1=5)=0$, which is very
different from the random distribution with bias 1.0026.

\begin{figure}
\centering
\includegraphics[width=7.6cm]{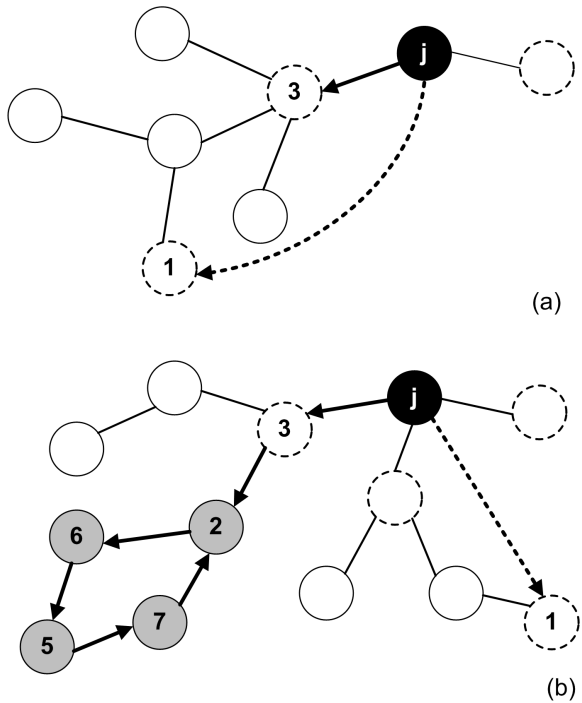}
\caption{Two possible schemes for changing the allele at locus $j$,
where nodes represent loci and the directed edge $j\to i$ represents that
the present allele at locus $j$ is $i$, while undirected edges are
irrelevant to the reassignment process. The black node is the node
(locus) $j$, the nodes with dashed border are the allele nodes in
this component, the gray ones are the influenced nodes and the
others are indifferent ones. (a) The new target node 1~(new allele)
is in the subgraph elicited from the node 3 (the present allele at
locus $j$). (b) The new target node 1 is not in the subgraph
elicited from the node 3.} \label{fig:reassignment}
\end{figure}

The idea behind the technique is well understood. Given a locus $j$,
we can replace the present allele with any other allele that
lies in the same component in a way that does not alter the
connectivity of the component hence causing no change in the
division encoded by the chromosome. To show its feasibility, we
focus on the component in which $j$ lies. Recall that a locus
represents a node and the allele at the locus represents the unique
neighbor the node adheres to. Consequently, the component is in the
form of a directed graph with unitary out-degree for each node.
There exist two possible schemes as shown in Fig.
\ref{fig:reassignment}. Note that the undirected edges are
irrelevant to the reassignment process; thus their
directions can be disregarded.

In the scheme depicted in Fig.~\ref{fig:reassignment}(a), we can
directly change the allele from 3 to 1 but still maintain the
connectivity of the component. For the scheme in
Fig.~\ref{fig:reassignment}(b), however, such direct altering of the
allele will split the original component. To deal with this case, we
study the travel in the component along directed edges, starting
from the node $j$. Since the subgraph elicited from node $j$ is
connected to the rest of the component through $j$, this travel
must end in a node that has passed. Let the path be $j\to x_1\to
x_2\to \cdots \to x_{k-1}\to x_k$. When $x_k\ne j$, we can
reestablish the connectivity by removing the last edge, reversing
the direction of each edge in the path, and adding a new edge
$x_1(3)\to j$. Note that the resultant graph meets with the
constraint that any node has only one outgoing edge. Therefore, we
can reset the alleles at those loci involved in the path. For
example, in Fig.~\ref{fig:reassignment}(b), the entire path is $j\to
3\to 2\to 6\to 5\to 7\to2$, so we can set the alleles according to
the path, $7 \to 5\to 6\to 2\to 3\to j$. Now, the allele at the
locus can be set to 1. As for the case $x_k=j$, we can directly alter
the alleles as in the scheme in Fig.~\ref{fig:reassignment}(a).

In the reassignment technique, we can also reassign the contribution
from the chromosome to the allele with the maximum label that lies
in the same component when performing locus statistics. More
generally, the method can also work as long as we arbitrarily specify a
fixed reassignment order for each locus, although different
prescriptions may produce different biases.

 Clearly, the reassignment technique is very useful
for community structure detection although it would not work when
applied to loci that have a single allele, i.e., the
corresponding nodes in the network are leaves. Moreover, this
special case can be readily eliminated by forbidding the mutation,
which may bring about the additional merit that it naturally reduces the
complexity of the problem. Since most real-world networks are
scale-free where substantial number of leaf nodes exist, this merit will be
very significant for finding communities in such networks.

\subsection{Population size}\label{subsec:popusize}
As in the MOGA, the unique parameter required to be provided in the MAGA is
the population size. The parameter may have significant influence
on the application of genetic algorithms. De Jong's experiment on a small suite of test functions showed~\cite{DeJong75} that the best population size was 50-100 for these
functions. There are also other empirical studies and theoretical
analyses of this parameter~\cite{Grefen86,Goldberg89}. In practice,
De Jong's setting has been widely adopted, which may be because this
choice gives a good tradeoff between the quality of the solution
and the cost of computation in many cases.

This popularity of the setting, however, does not exclude the
development of genetic algorithms working with a variable population
size. A few examples of the class of algorithms can be found in
\cite{Arabas94,Affen07,Hu10}. Although one of these mechanisms may
be beneficial to be incorporated into MAGA, in this work we does not
take it into account.

Since we expect that all alleles at a locus can simultaneously appear in
the population, a population size that is
greater than the degrees of most nodes in the network would be preferable . As mentioned
before, most real-world networks are scale-free, so the degrees of
most nodes in these networks are smaller than 50. Considering this fact
and the cost of large population size, we would like to take a fixed
value from the interval between 50 and 200.

\subsection{Convergence and its speeding up}\label{subsec:covergence}

The MOGA was reported to perform well in the application to solve the
knapsack problem~\cite{Szeto06}, where all the loci have two
alleles, 0 and 1. For many cases, however, its performance will be
hindered by two factors. One factor is the misguiding by the
allele standard deviation mentioned above. The other is that in
the evolution of each generation the fittest individual(s) actually
will not participate in the mutation unless others supersede
it~(them).

In fact, despite fulfilling the elite preservation
strategy~\cite{Bean94,footnote1} that assures convergence for
a SGA toward the global optimum, the MOGA does not guarantee such
convergence and even may end with a nonlocal optimum solution.
Consider a case where the $N_P-1$ fittest chromosomes have identical
fitness and the remaining one has a lower fitness value. Those fittest
should be passed to the next generation while the remaining one will
mutate with very high probability. If the mutation happens to
produce a chromosome with the same fitness as the others, this will
unexpectedly terminate the evolutionary process.

Moreover, it is helpful to notice that the present fittest chromosomes, if not a local
optimum, always can perform a local search to reach a local optimum.
Consequently, it is preferable to modify the rule for mutation so as
to allow for local search, which refers to performing a random mutation
on a single locus. The mutation operation is powerful in that it may
lead to a node moving between different components, a component splitting
or two components merging.

Interestingly, we found that in many cases it may be useful for the
local search to perform a special splitting operation with a low
probability~(for example, 0.1). The splitting operation on a
component drawn randomly can be implemented by a bipartitioning in
the spectral method~\cite{Newman06a,Newman06b,footsplit}. Let the
number of edges in the component be $M_c$. For the power method, it
needs $O(N)$ multiplications to converge the lead vector of a matrix
of size $N$, which leads to a run time $O(N^2)$ for a bipartition in
the spectral method~\cite{Newman06a}. In order to not increase the time
complexity of each generation's evolution, the multiplication is
executed at most $N \log N /M_c$ times.

Combining the above considerations, the overall procedure of the MAGA for
community detection can be described as follows.

(1)~The connectivity of the network of interest is fed into the MAGA.
The algorithm then creates $N_P$ initial feasible solutions, each
locus of which is initiated with a random allele.

(2)~At each generation, the MAGA first duplicates 10\% the fittest
chromosomes of previous generation for the current generation.

(3)~The MAGA then reproduces $0.9N_P$ individuals by selecting from the
previous generation in proportion to their fitness to prepare for
mutation.

(4)~The fitness and the fitness cumulative probability for
chromosomes are evaluated using Eqs.~(\ref{eq:fitness})
and~(\ref{eq:cumufitness}), respectively; immediately, the bias for each locus using Eqs.~(\ref{eq:bias})
and~(\ref{eq:allele-newdistr}) is evaluated, and then these loci are ranked according
to their biases.

(5)~The individuals reproduced in step 3 are swept, and the
chromosome $i$ selected with the same probability $1-C(f(i))$ as in the MOGA; if
the chromosome is chosen then the mutation aforementioned is performed, otherwise
a local search for the fitter individuals is performed.

(6)~Steps 2--5 are repeated until a certain termination criterion has
been met. Otherwise, the MAGA outputs the best partition with the
highest fitness.

Since in step 3 the fitter individuals incline to be reproduced because of their higher fitness, step 4 enables the reproduced fittest
individuals always to perform a local search. Step 2 maintains the elite preservation strategy in case of the destruction of the strategy in step 5. In this way, the MAGA not only can converge to the global optima, also can speed up the process.

The most time-consuming operations in each generation are evaluating
the bias and fitness with $O(M)$ time, and ranking the loci with
$O(N\log N)$ time. This ranking operation has seemingly slightly higher
complexity than an $O(M)$ operation if the network is sparse. In
fact, it can be performed faster than those operations with $O(M)$ time
since the latter need to be repeated $N_P$ times. Therefore, the overall
time cost for each generation of the MAGA is $O(M)$ like that of SGAs.

\section{Results}\label{sec:results}

In this section, we empirically study the effectiveness of the MAGA by
applying it to model bipartite networks and several real bipartite
networks. In both cases, we show that MAGA is superior to SGAs and the MOGA~\cite{footnote2}, and it also can compete with the nice BRIM
~(bipartite, recursively induced modules)~\cite{Barbe07} algorithm that dedicated to bipartite networks. We also tested the
performance on several real unipartite networks, comparing with
several well-known methods for unipartite networks in the
literature.

\subsection{Model bipartite networks}

To test how well our algorithm performs, we have applied it to
model bipartite networks with a known community structure. A model
network can be constructed in two steps. The first step is to
determine the layout of nodes in the network, i.e., to specify the
number of communities $N_C$, and the numbers of nodes of two types
included in each community $N_A$ and $N_B$, as well as to assign
group membership to these nodes. Next, the dispersion of
edges is determined by specifying the intracommunity and intercommunity link
probabilities $p_{\rm in}$ and $p_{\rm out}$, such that $p_{\rm
in}\ge p_{\rm out}$.

For simplicity, all communities assume the same values of $N_A$ and
$N_B$. We set $N_C=5$, $N_A=12$, and $N_B=8$ as used
in~\cite{Barbe07}. One might expect that as $p_{\rm in}$ is markedly
greater than $p_{\rm out}$ the networks exemplifying the model
have significant community structure that tends to be detected.
Conversely, as $p_{\rm out}$ approaches $p_{\rm in}$, the network
examples become more uniform and their modular structure
becomes more obscure. In this experiment, $p_{\rm in}$ is fixed
at the value of 0.9 while $p_{\rm out}$ is varied by tuning $p_{\rm
out}/p_{\rm in}$ from 0.1 to 0.9 with steps of 0.1. We have tested on
such models the performance of the MAGA as well as of the SGA and the MOGA, each
exemplified with ten networks. On each example we ran these
algorithms ten times.

For evaluating the quality of solutions, both the modularity and
the normalized mutual information~(NMI)~\cite{Danon05} are useful. But the NMI is more
suitable for the current case since the optimal~(correct) division of
the model network is known in advance. This measure takes its
maximum value of 1 when the found division perfectly matches with the
known division while it takes 0, the minimum value, when they are
totally independent of each other. Accordingly, we employed the stop
criterion that the algorithms reach the predefined generation
size~(maximum number of generations) or the NMI reaches its maximum
value.

\begin{figure}
\centering
\includegraphics[scale=1]{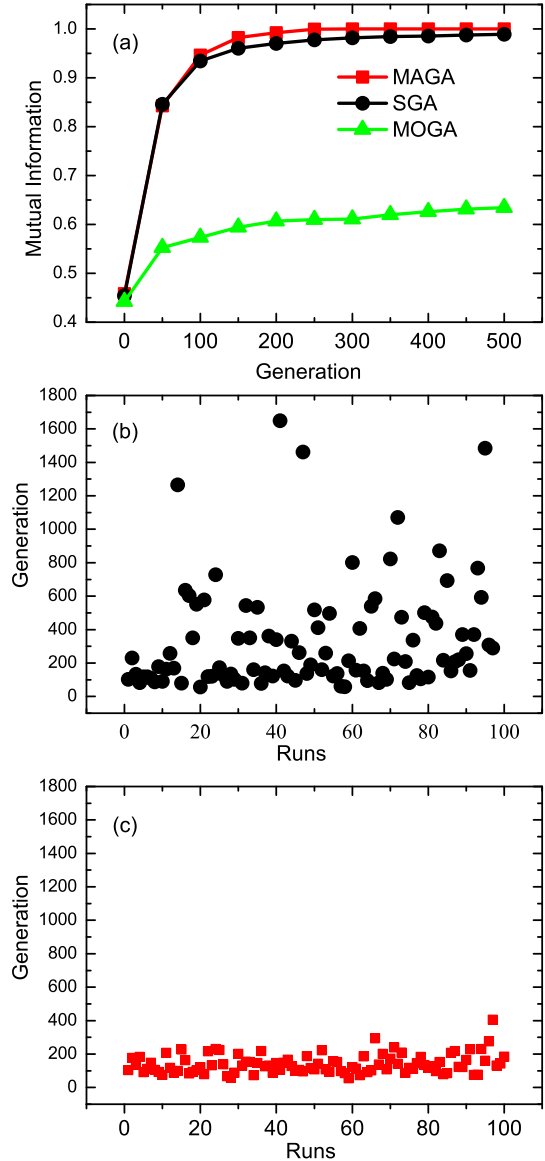}
\caption{(Color online)~Performance on bipartite model networks with
$p_{\rm in}=0.9$ and $p_{\rm out}/p_{\rm in}=0.1$. The generation
size is set to 2000. (a) Variation of normalized mutual information
over first 500 generations. (b) Distribution of the number of
generations needed to reach the optimum using the SGA. More than half the number of
generations are over 200. (c) Distribution of the number of
generations needer to reach the optimum using the MAGA. There are 83 runs in which
the number of generations is less than 200.}
\label{fig:model1}
\end{figure}

\begin{figure}
\centering
\includegraphics[scale=1]{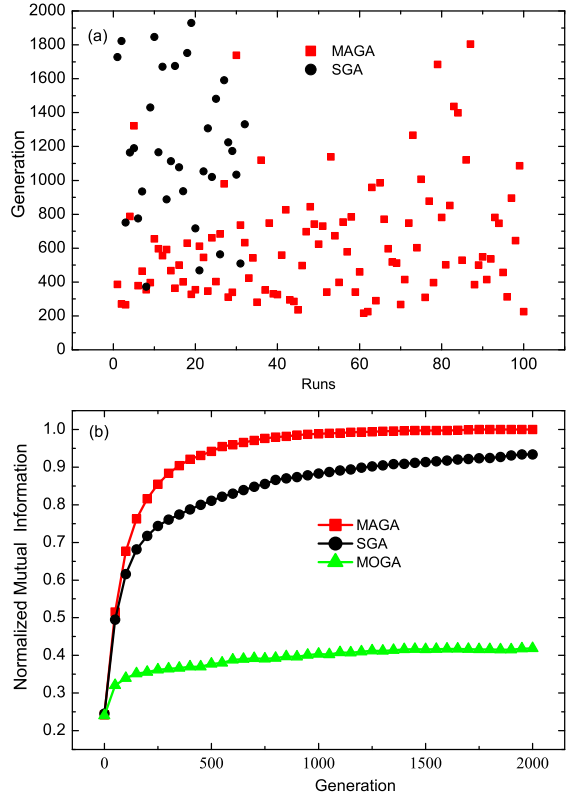}
\caption{(Color online)~Performance on model networks with $p_{\rm
in}=0.9$ and $p_{\rm out}/p_{\rm in}=0.2$. The generation size is
set to 2000. (a) Distributions of the number of generations to reach
the optimum using the SGA and MAGA. There are 32 black circle points and 100
red box points respectively representing the number of generations
needed to reach the optimum using the SGA and MAGA. Most numbers of
generations for the SGA are distributed above 1000 while for the MAGA most
are below 800. (b) Variation of normalized mutual information with
the number of generations. Each point is the average over the 100
runs.} \label{fig:model2}
\end{figure}

Figures~\ref{fig:model1} and~\ref{fig:model2} display the
performance comparison between such genetic algorithms for $p_{\rm
out}/p_{\rm in}=0.1$ and $p_{\rm out}/p_{\rm in}=0.2$, respectively.
The generation size is set to 2000. For both cases, the MAGA and the SGA
remarkably outperform the MOGA. From Fig.~\ref{fig:model1}~(a), we can
see that the the MAGA is appreciably faster than the SGA, although both perform
well since the mutual information rapidly exceeds 0.9. In our test,
each run of MAGA on all ten example networks consistently gave
the optimal division, i.e., produced 100 numbers of generations
less than 2000. For the SGA, 97 runs gave the optimum division. Their
distributions of the number of generations needed to reach the optimum,
reported in Figs.~\ref{fig:model1}(b) and \ref{fig:model1}(c) further reveal their
difference in speed (in terms of the number of generations).

When $p_{\rm out}/p_{\rm in}=0.2$, it is more difficult to identify their
community structure of the example networks relative to the previous ones. The SGA succeeded in obtaining the optimum division
in 32 runs. In sharp contrast, each run of the MAGA gave the optimum
division. More information on the distributions of the number of
generations is provided in Fig.~\ref{fig:model2}~(a). Also, in
Fig.~\ref{fig:model2}(b), the variations of the mutual information with regard to
the SGA and the MAGA illuminate that there exists a greater performance
difference between them than in the case of $p_{\rm out}/p_{\rm
in}=0.1$.

Even for $p_{\rm out}/p_{\rm in}=0.1$, the MOGA was not observed to reach the optimum
solution in its first 2000 generations was not observed. Actually, the
MOGA performed so poorly that it was even much slowly than the SGA
as shown in Figs.~\ref{fig:model1}~(a) and ~\ref{fig:model2}~(b). We argue that the main reason
for this is that the use of an incorrect informative measure for the loci
has misguided the algorithm.

\begin{figure}
\includegraphics[scale=0.8]{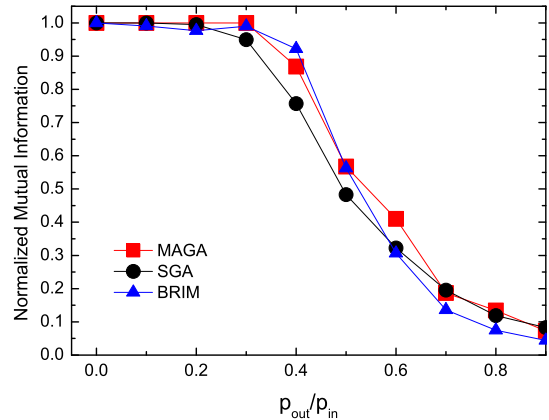}
\caption{(Color online)~Variation of performance of the algorithms
with different $p_{\rm out}/p_{\rm in}$. Each point is the average
over ten sample networks. For $p_{\rm out}/p_{\rm in}=0.1$ and
$0.2$, the generation size is set to 2000; for other values, the
size is set to 3000.} \label{fig:AllModel}
\end{figure}

We have made a more extensive performance comparison. Figure ~\label{fig:AllModel}
shows the variations of accuracy of the MAGA and SGA as well as BRIM
against changes of $p_{\rm out}/p_{\rm in}$. For the model
networks, assigning each of the nodes from the smaller groups to its
own module is a better strategy for BRIM that
will lead to a precise division. To be fair~\cite{footnote3}, we
picked the best division from the ten runs on each sample
network and then averaged over ten examples for a particular
$p_{\rm out}/p_{\rm in}$.

\subsection{Southern women network}

As the first example of a real bipartite network, we study the
southern women network~\cite{Davis41}. The social network consists
of 18 women and 14 events for which the data were collected by Davis
\emph{et al.} in the 1930s, describing the participation of the women in these events.
It has been extensively used as a typical instance for investigating
the problem of finding cohesive groups hidden in social networks;
see Ref.~\cite{Freeman03} for a useful review.

We have performed the MAGA ten times on this network, with the population
size 100 and the generation size 3000. Unlike the BRIM algorithm for
which initial state is important, initial states are generally irrelevant~(or weaker relevant) for genetic algorithms to they can succeed
in finding a quite good solution. For each run, the MAGA found
the best solution so far, with $Q=0.3455$.

\begin{figure}
\includegraphics[scale=0.32]{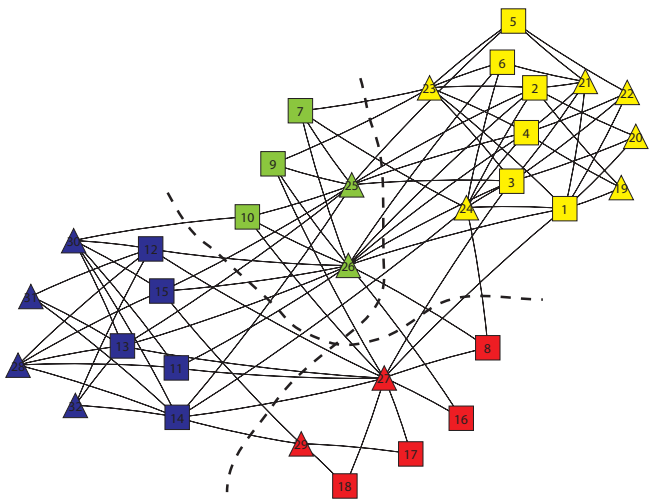}
\caption{(Color online)~Southern women network~(dashed lines indicate the division found by the MAGA). Each community consists of those nodes with the same color~(level of scale),
including women and events represented by box and triangle events
respectively. } \label{fig:women}
\end{figure}

Figure~\ref{fig:women} shows the community structure identified in
the southern women network using the MAGA. This division is exactly the
same as that found with BRIM with the initial strategy that
begins with assigning all events to a single community. We have also applied the SGA and the MOGA to this network with the same population size
and generation size. A simple performance comparison between
them is shown in Table \ref{tab:women1}, which lists the success
times for reaching the best solution, the minimum~(MinGen) and maximum number of generations~(MaxGen) to reach the optimum, and the average normalized mutual
information~($I^*_{\rm norm}$), and average modularity~($Q^*$).

No matter what we are concerned about, the speed or the quality,
the MAGA again has an evident advantage over the SGA and MOGA. Table
\ref{tab:women2} shows the accuracy of the MAGA in comparison with other
methods.

\begin{table}[h]
\caption{Performance comparison between the SGA, MOGA and MAGA on the southern women network. Each algorithm runs ten times. Here, success means that the algorithms
have found the best solution before they reaches the generation size 3000.}
\label{tab:women1}
\begin{ruledtabular}
\begin{tabular}{lddddd}
    Method & \multicolumn{1}{c}{Succ.} &\multicolumn{1}{c}{MinGen.}&\multicolumn{1}{c}{MaxGen.} &\multicolumn{1}{c}{$I^*_{\rm norm}$} & \multicolumn{1}{c}{$Q^*$} \\ \hline
    SGA &4&2011&2924 &0.8923 &0.3454\\
    MOGA &0 &-  &- &0.7997 &0.3448\\
    MAGA &10&87&1830 &1 &0.3455\\
\end{tabular}
\end{ruledtabular}
\end{table}

Most previous studies assigned these women to groups
depending on their interests. Davis \emph{et al.}~\cite{Davis41} assigned the women
to two groups, labeled 1-9 and 9-18. Woman 9 can be considered as an
overlapping node of the two groups in a sense, but should be
exclusively included in one group by the currently used community
definition. We may label the division with 9 and 1-8 in the same
group as ``Davis 1", and the alternative division~(9 is grouped
with 10-18) as ``Davis 2".

Doreian \emph{et al.}~\cite{Doreian04} took the definition of a bipartite
community composed of two types of nodes and proposed several
divisions, with the accuracy of the division with the highest
modularity shown in Table~\ref{tab:women2}. We call the BRIM algorithm
using the strategy of (1) assigning all events to a single module and (2)
assigning each event to its own module ``BRIM 1" and ``BRIM 2," respectively. Barber~\cite{Barbe07} reported its accuracy when using
such strategies on the network; these results also can be found in
Table~\ref{tab:women2}.

\begin{table}[h]
\caption{Performance comparison on the southern women network, where
some data are drawn from \cite{Barbe07}.} \label{tab:women2}
\begin{ruledtabular}
\begin{tabular}{lddd}
    Method & \multicolumn{1}{c}{Communities}& \multicolumn{1}{c}{$Q^*$} & \multicolumn{1}{c}{$I^*_{\rm norm}$} \\ \hline
    MAGA & 4&0.3455 &1 \\
    BRIM 1 & 4&0.3455 &1 \\
    BRIM 2 & 2 &0.3212  &0.5803\\
    Davis 1 &2  &0.3106 &0.4466\\
    Davis 2 &2  &0.3184 &0.4513\\
    Doreian &3  &0.2939 &0.6077\\
\end{tabular}
\end{ruledtabular}
\end{table}

\subsection{Scotland corporate interlock network}

The second real-world bipartite network we have used as a test on is
the Scotland corporate interlock network~\cite{Scott80}. This network
describes the corporate interlock pattern between 136 directors and
the 108 largest joint stock companies during 1904-1905. As it is
disconnected, we focus merely on its largest component, which
comprises 131 directors and 86 firms. In the following, the word
``network'' consistently indicates this component.

The BRIM algorithm found poorer divisions of this network with
$Q=0.5663$ and $Q=0.3987$, using the strategies of assigning all
directors to unique modules or to the same module. With the adaptive
binary search technique, the BRIM algorithm, when using the strategy of
randomly assigning directors to modules, may find a much
better solution with $Q=0.663$($\pm$0.002). Based on the
experimental results, the author of \cite{Barbe07} suggested that the network comprise
approximately 20 communities.

\begin{figure}
\includegraphics[scale=0.8]{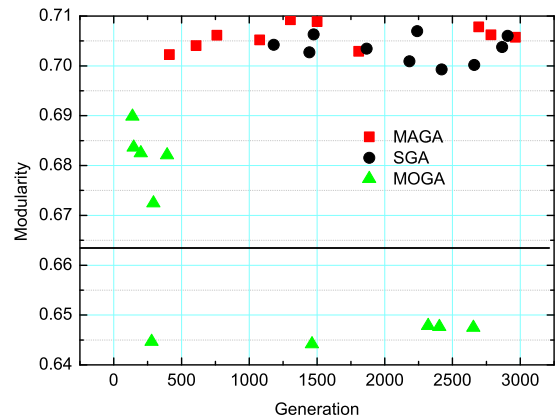}
\caption{(Color online)~Distributions of the divisions returned by
the SGA, MOGA and MAGA. The black horizontal line indicates the best
bipartite modularity reported in \cite{Barbe07} using the BRIM
algorithm.} \label{fig:scotland_distribution}
\end{figure}

Similarly, we have examined the performance of three algorithms
on this network by running ten times with the same settings as
before. Figure \ref{fig:scotland_distribution} shows the
distributions of the solutions returned by SGA, MOGA, and MAGA.
Obviously, both the SGA and MAGA definitely exhibit higher accuracy than
BRIM and the MOGA.

Moreover, the MAGA appears preferable to the SGA. In the experiment,
the modularity of the best division found by the SGA, $Q =0.7070$, is
less than those of the best two divisions~($\pi_1$ and $\pi_2$) found by
the MAGA with $Q=0.7093$ and $Q=0.7089$. On the other hand, as shown in
Fig.~\ref{fig:scotland_distribution}, for the MAGA most of the ten divisions
including the best two are found during the first 2000 generations
while for the SGA six of the ten divisions are found after 2000 generations.

\begin{figure*}
\includegraphics[scale=0.4]{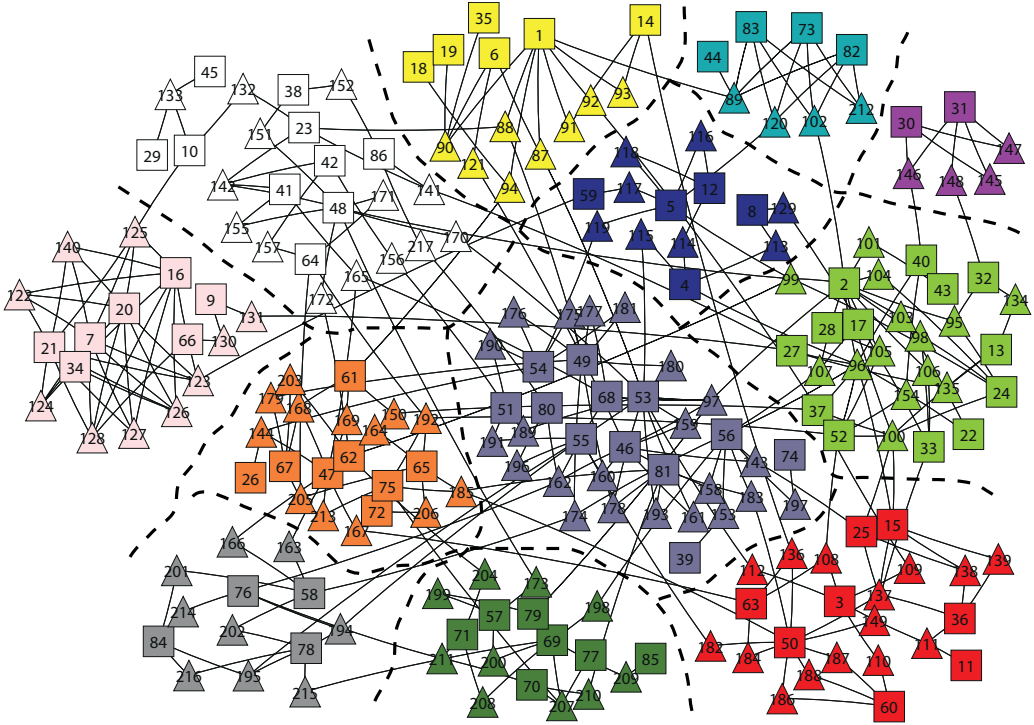}
\caption{(Color online)~Scotland corporate interlock network~(dashed lines indicate the division found by the MAGA). Each community consists of those nodes
with the same color(level of scale), including firms and directors represented by
boxes and triangles respectively. } \label{fig:scotland_community}
\end{figure*}

In closing, we would like to give a simple evaluation of the
reliability of the solutions. We calculated the normalized mutation
information between any pairs of solutions returned by the MAGA. The
maximum value of the NMI is between $\pi_1$ and $\pi_2$ and is equal
to 0.9191, indicating that they are very similar. Simultaneously, for
each solution, we calculated the average of the NMI between that
division and other divisions. We found that $\pi_2$ has the largest
value, 0.8459, and $\pi_1$ has the third largest value, 0.8248. These
facts lend confidence in the reliability of the optimum divisions
obtained, $\pi_1$ and $\pi_2$. Figure \ref{fig:scotland_community}
shows the community structure of this network according to  $\pi_2$.
Clearly, the MAGA indeed has given a very accurate division of this
network.

\subsection{Unipartite networks}\label{sec:unipartite}

The MAGA can also be applied to unipartite networks by optimizing the
bipartite modularity after the transformation as mentioned in
Sec.~\ref{sec:modularity}. Being a kind of genetic algorithm,
however, the MAGA can directly optimize the unipartite modularity as the SGA
does~\cite{Pizzuti08}, which distinguishes it from certain
methods such as the SR method which is required to develop different
versions for different classes of
networks~\cite{Newman06a,Newman06b,Leicht08,Barbe07}. Furthermore,
the modularity consistency revealed in Sec.~\ref{sec:modularity}
means that the MAGA can also more effectively optimize unipartite
modularity so that we only focus on the comparison with those
well-known methods, including the Girvan-Newman (GN)
algorithm~\cite{Girvan02}, EO~\cite{Danon05},
SR~\cite{Newman06a,Newman06b}, and SA~\cite{Guimera05}.

To test the performance of the MAGA on unipartite networks, we have
considered several real networks with different scales: the Zachary
karate club network~\cite{zachary}, the jazz musicians
network~\cite{Gleiser03}~(Jazz), the Caenorhabditis elegans metabolic
network~\cite{Jeong00}~(C. elegans), the email network of University Rovira i
Virgili \cite{Guimera03}~(Email), a trust network of users of the
Pretty-Good-Privacy~(PGP) algorithm for information security
\cite{Guardiola02}, and a coauthorship network of scientists working
in condensed matter physics~\cite{Newman01a}~(Cond-mat).

The EO and SR methods clearly outperform the original method for
detecting communities~(the GN method); they may both be viewed as the
representatives of modularity maximization approaches in that they
can achieve a good tradeoff between speed and accuracy. As shown in
Table~\ref{tab:unipartite}, the MAGA almost consistently outperforms the EO
and SR methods for these networks. Interestingly, for the Zachary
network the MAGA found the accurate solution with
$Q=0.4198$~\cite{Agarwal08,Lai10}, while neither the EO nor SR nethod can find it in
spite of the fact that the network is very simple. Furthermore, for
the larger networks the gap in performance tends to widen; for
example, the maximum modularity difference approaches 18\% (11\%)
relative to the EO~(SR) method for the largest network studied.

Even when compared with SA method, which is widely considered as the most accurate
modularity maximization method, the MAGA may give a higher modularity while
significantly reduce the time cost. In fact, the SA method theoretically allows finding the global optima of modularity, but the exponential complexity restricts it only to finding a better local optimum and to resolving the network of scale only up to $10^4$. The performance of the SA listed in Table~\ref{tab:unipartite} was reported in running on an Intel PC
with two 3.2 GHz processors in~\cite{Agarwal08}, wherein the authors
proposed an accurate method that can be competitive with the SA method but has
very high memory demand. We ran the MAGA ten times for all the networks,
with predefined generation size, on an Intel PC with two
2.93 GHz processors. The last two columns of Table~\ref{tab:unipartite} shows the number of generations and the running time needed to find the maximum modularity in the runs.

\begin{table*}
\caption{Performance comparison of the MAGA, Girvan-Newman~(GN), extremal optimization~(EO),
spectral relaxation~(SR), and simulated annealing~(SA) methods in terms of
modularity and running time (only for the SA and MAGA) for unipartite
networks. The modularity in bold font represents the maximum modularity obtained for the network, with the corresponding number of generations and time shown in the last two columns. The running time for the SA or MAGA is measured in minutes~(min) or seconds~(s).}
\label{tab:unipartite}
\begin{ruledtabular}
\begin{tabular}{lrcrcrrcccr}
             &      &       &       &       & \multicolumn{2}{c}{SA(0.999)} & \multicolumn{4}{c}{MAGA} \\
             \cline{6-7} \cline{8-11}
    Network  &Size  & GN    & EO    & SR     & $Q$   & Time    & GenSize & $Q$     &Generations & Time \\
\hline
    Zachary   & 34   & 0.401 & 0.419 & 0.419 &\textbf{0.420} &12s    & 100   & \textbf{0.420} &5     &0.1s\\
    Jazz      & 198  & 0.405 & \textbf{0.445} & 0.442 & \textbf{0.445} &58min    & 8000  & \textbf{0.445} &7222  &19min\\
    C.elegans & 453  & 0.403 & 0.434 & 0.435 & 0.450 &146min   & 8000  &\textbf{0.452} &3487  &12min\\
    E-mail    &1133  & 0.532 & 0.574 & 0.572 & 0.579 &1143min  & 10000 & \textbf{0.581} &9280  &72min\\
    PGP       &10680 & 0.816 & 0.846 & 0.855 &  -    &-      & 20000 & \textbf{0.881} &19867 &610min\\
    Cond-mat  &27519 & -     & 0.679 & 0.723 &  -    &-      & 30000 & \textbf{0.802} &29995 &3517min\\
\end{tabular}
\end{ruledtabular}
\end{table*}

\section{Conclusion}\label{sec:conclusion}

We have shown both that unipartite and directed networks can be
equivalently represented as bipartite networks, and their modularity
is just the corresponding bipartite modularity. This implies that
bipartite networks can be considered as an extensive class of
networks including unipartite and directed networks, and that
detecting communities in bipartite networks provides a uniform
framework for solving the problem in various networks. Therefore,
methods for detecting community structure of bipartite networks
generally can be applied to unipartite and directed networks.

We have presented an adaptive genetic algorithm, the MAGA, for the task
of community structure detection. This algorithm is based on the MOGA
which was presented with the aim of improving the performance of
traditional genetic algorithms. But we have shown that the MOGA has a
poor performance as applied to this task. In fact, we have revealed
the MOGA would be misguided by the allele standard deviation and does not guarantee the convergence to global optima. In the MAGA,
we introduced a different measure for the informativeness of loci, a
modified rule for mutation and a reassignment technique. These
ingredients jointly make the MAGA more effectively optimize objective
function for community structure detection. The experiments on
bipartite (model and real) networks have consistently shown that
the MAGA outperforms the MOGA, SGA, and BRIM. Compared to BRIM, another
advantage is that the MAGA can automatically determine the number of
communities. The results on unipartite networks indicate that
the global optimization method is indeed more accurate than the EO and
SR methods as expected, and that it also can attain or even
outperform the accuracy of the SA method in a significantly shorter time,
which is crucial for analyzing large networks.

The time complexity of each generation evolution of the MAGA is $O(M)$,
and the overall time demand of this algorithm depends on the population
size and the generation size~\cite{Footgensize}. Although the MAGA can
theoretically find the global optima of an objective function, the
quality of solutions delivered by the MAGA rests in practice on the
generation size given the population size. Owing to the lower
complexity of each generation evolution, we can run enough generations to get a high-quality solution. Empirical results showed
that the MAGA can effectively resolve the community structure of networks at
many scales up to $10^5$, which have covered many kinds of real
networks such as social, metabolic, and technology
networks. Beyond these scales there are several nice local methods
available \cite{Lancichinetti09,Raghavan07,Leung09,Ronhovde10},
while the performance of our algorithm on networks with such scales
needs to be further explored. On the other hand, since a parallel
implementation of the MAGA allows each of the most time-consuming
operations on $N_P$ chromosomes to be simultaneously calculated by
assigning them to multiprocessors of a highly efficient computer, it
seems that even for networks of millions of nodes the MAGA is still a
promising method for accurate detection of their community structure.

Methodologically, the MAGA for community detection is based on the idea
of optimization. So the accuracy is determined by the selection of
an objective function. Here, we use the (bipartite) modularity as the
object to optimize, which certainly may suffer from the resolution
problem although this may not be severe for many real networks. On the
one hand, the resolution problem essentially is favorable for gaining
deeper insight into the structure of networks~\cite{Arenas07}. On the
other hand, the effect of this problem may be circumvented or
alleviated as needed. For example, the MAGA can perform network
preprocessing with random walk~\cite{Lai10} before optimizing or take
an alternative objective function~\cite{Rosvall08} instead of
the modularity. Also we can combine several high-quality solutions to
obtain a more accurate division of the network of
interest~\cite{Raghavan07,Pardo07}.

Overall, the MAGA enables us to accurately and effectively detect community structure
for various networks including bipartite, unipartite,
directed, and weighted networks so long as it takes the corresponding modularity as the fitness function. The evolutionary method can return multiple high-quality
solutions with no bias, which may provide some useful information on
the reliability of the solutions of interest and may be combined in
a way to obtain a better solution. Finally, we believe that as an
effective discrete optimization method~(the special reassignment
technique can be switched off as needed) it will find more
applications in many fields.

\begin{acknowledgments}
This research was supported by the National Basic Research Program
of China under Grant No. 2007CB310806, the National Natural Science
Foundation of China under Grants No. 61074119, No.~60873040 and
No.~60873070, Shanghai Leading Academic Discipline Project No. B114.
J.-H. G. was also supported by the Shanghai Education Development Foundation under Grant No. 09SG23.
\end{acknowledgments}


\begin{references}
\bibitem{footnote1}From the discussion in Sec.~\ref{subsec:moga}, the MOGA will duplicate the
fittest individual(s) to a next generation.
\bibitem{footsplit}After the bipartitioning, we need to update the chromosome by encoding the two resulting groups of nodes. This can be implemented in time of $O(m)$ by the breadth-first search algorithm~\cite{Aho83}, where $m$ is the total number of edges of these nodes involved.
\bibitem{footnote2}Since the splitting operation can be incorporated
into the SGA in a certain way, to make a fair comparison between the MAGA and
SGA, all the results on bipartite networks we reported do not
involve the operation. For all the tested unipartite networks the
operation is employed and may improve the efficacy of the MAGA, even if it
often yields a rough bipartition.
\bibitem{footnote3}Picking the best one
of the run on each example is reasonable because the
population size is moderate so that it can be increased and hence
improve the accuracy of each run; that is, we adopted a better
strategy as is done in BRIM.
\bibitem{Footgensize}A simple strategy for selection of the generation size consist of two steps: (1)~Run the algorithm with an empirical value of generation size 5-10 times. For smaller networks the generation size can take a value as large as ten times of the network size, while for larger networks a value equal to the network size can be used. (2)~Appropriately increase the generation size according to the results and the average time cost if the MAGA does not converge. If these results are very different and the time cost is low, this may indicate that the generation size should be mediated with a large increment; otherwise, it may need a small increment. Generally, this adjustment should be made at most twice or three times.
\end{references}
\end{document}